\documentclass[11pt,a4paper]{article}
\pdfoutput=1

\usepackage{jcappub}
\usepackage[margin=0.95in]{geometry}

\usepackage[utf8]{inputenc}
\usepackage{graphicx}
\usepackage{verbatim}
\usepackage{epsfig}
\usepackage{subfigure}
\usepackage{color}
\usepackage{multirow}
\usepackage{comment}
\usepackage{slashed}
\usepackage{amsmath}
\usepackage{ulem}
\usepackage{hyperref}
\newcommand{\arXiv}[2]{\href{http://arxiv.org/pdf/#1}{{\tt [#2/#1]}}}
\newcommand{\arXivold}[1]{\href{http://arxiv.org/pdf/#1}{{\tt [#1]}}}

\usepackage{amssymb,amsmath,amsfonts,mathrsfs}
\usepackage[dvipsnames]{xcolor}
\usepackage{graphicx}
\usepackage{longtable}
\usepackage{verbatim}
\usepackage{color}
\usepackage[mathscr]{euscript}
\usepackage{framed}
\usepackage{mdframed}  

\usepackage{cancel}
\usepackage{color}
\usepackage{hyperref}
\hypersetup{colorlinks, citecolor=bluscuro, linkcolor=black, urlcolor=bluscuro}
\definecolor{rossos}{cmyk}{0,1,1,0.55}
\definecolor{bluscuro}{rgb}{0.15, 0.2, .85}
\definecolor{bluchiaro}{cmyk}{1,.3,0.,0.1}
\graphicspath{{./Figures/}}

\numberwithin{equation}{section}
\renewcommand\theequation{\arabic{section}.\arabic{equation}}

\usepackage{tikz}

\usepackage{tcolorbox}

\newcommand{\nn}{\nonumber}

\renewcommand{\r}{r_*}

\newcommand{\dd}{{\rm d}}

\newcommand{\be}{\begin{equation}\begin{aligned}}
\newcommand{\ee}{\end{aligned}\end{equation}}

\newcommand{\bbe}{\begin{align}}
\newcommand{\eee}{\end{align}}

\newcommand{\bea}{\begin{eqnarray}}
\newcommand{\eea}{\end{eqnarray}}

\def\beq{\begin{equation}}
\def\eeq{\end{equation}}

\def\beqa{\begin{eqnarray}}

	\def\eeqa{\end{eqnarray}}
	
\def\lsim{\mathrel{\rlap{\lower4pt\hbox{\hskip0.5pt$\sim$}}
		\raise1pt\hbox{$<$}}}         
\def\gsim{\mathrel{\rlap{\lower4pt\hbox{\hskip0.5pt$\sim$}}
		\raise1pt\hbox{$>$}}}         

\def\eeqa{\end{eqnarray}}

\def\bq{\begin{quote}}
\def\eq{\end{quote}}

 at 10truept

\title{ \huge    Quasinormal Modes and Love Numbers \\ of   Kerr Black Holes 
 from AdS$_{\mathbf 2}$ Black Holes}

\author[a]{A. Kehagias,}
\author[b,c]{D. Perrone,}
\author[b,c]{A. Riotto}

\affiliation[a]{Physics Division, National Technical University of Athens, Athens, 15780, Greece}
\affiliation[b]{
	Department of Theoretical Physics and Center for Astroparticle Physics (CAP) \\
			24 quai E. Ansermet, CH-1211 Geneva 4, Switzerland}
\affiliation[c]{Gravitational Wave Science Center (GWSC), Universit\'e de Gen\`eve, CH-1211 Geneva, Switzerland}

\abstract{We show that the linear perturbations of any spin field in the near-zone limit of the Kerr black hole are identical to those of an AdS$_2$ black hole which enjoys the same basic properties of the Kerr black hole. Thanks to this identification, we  calculate the spectrum of the 
quasinormal modes and the Love numbers of  Kerr black holes using an  AdS$_2$/CFT$_1$ correspondence and a  group theoretical approach.
\vskip 2cm
\noindent

}

\emailAdd{kehagias@central.ntua.gr}
\emailAdd{davide.perrone@unige.ch}
\emailAdd{antonio.riotto@unige.ch}

\begin{document}

\maketitle
\flushbottom

\newpage

\section{Introduction}
\label{sec:intro}
Linearized perturbations of Black Holes (BHs)  in General Relativity has a long history \cite{ch}.  A fundamental  role is played by the concept of QuasiNormal Modes (QNMs) \cite{rev,br, pani}. They are oscillations with a  purely ingoing wave condition at the BH horizon and purely outgoing wave condition at large distances. They are relevant as they  provide the late-time evolution of linear perturbations in the BH exterior. In particular, in the merger of two BHs, during the  final stage (the so-called ringdown),  the QNMs dominate the BH  response to any kind of disturbance. Their frequencies are uniquely determined by the BH mass, spin and charge, thus  providing a unique tool  to study BHs whose  spectroscopy  is a major goal of gravitational wave astronomy. 
QNMs are also relevant as they may provide a link between classical General Relativity and Quantum Gravity. The static response of BHs to external disturbances, captured by the so-called Love numbers \cite{mm}, has also  recently attracted significant attention both from the observational and theoretical sides \cite{hind,poi}.

Motivated by these general considerations, in this paper we investigate the 
 QNMs and the Love numbers of Kerr BHs.  We will exploit  the symmetry properties of the Kerr BH in the so-called near-zone region where the Kerr BH metric can be effectively described by the  simpler geometry of an AdS$_2\times S^2$. The role of other symmetries has been highlighted in describing properties of Kerr BHs in the past. The 
 near-horizon extreme Kerr  geometry has an ${{\rm SL}(2,\mathbb{R})}\times U(1)$ isometry group \cite{bh} and it has been  conjectured that quantum gravity in the near-horizon extreme Kerr BH geometry with certain boundary conditions is equivalent to a chiral conformal field theory in $(1+1)$ dimensions \cite{kcft}. More recently, a 
correspondence between gravitational perturbations and quantum Seiberg-Witten curves, as well as conformal field theory and localization techniques,  have been used to compute the spectrum of QNMs of asymptotically flat Kerr Newman BHs \cite{sw1,sw2,sw3,sw4}. Symmetry arguments have also been invoked to explain the vanishing of the so-called Love number, characterizing the BH static and non-dissipative tidal response \cite{love1,love2,love3}. 
 
 Extending the recent results of Ref. \cite{Hui} valid for a massless scalar field, we will show that the identification of the Kerr BH with an AdS$_2$ BH in the near-zone limit holds at the linear level for any spin field. We will prove that near-zone geometry is indeed that of an AdS$_2$ BH possessing a  horizon and an ergosphere  sharing  the same  properties of those of  the Kerr BH.  Thanks to such an identification, we will calculate the spectrum of the  QNMs and the Love numbers of Kerr BHs making use of  AdS$_2$/CFT$_1$ correspondence and group theory arguments\footnote{For alternative analytical ways to calculate the spectrum of  QNMs in the extreme Kerr BHs, see for instance Refs. \cite{other1,other2,other3,other4,other5,other6}.}.

 The paper is organized as follows. In section 2 we summarize the Kerr geometry, while section 3 is devoted to the description of the near-zone geometry. In section 4 the horizon and the ergosphere of the AdS$_2$ BH are discussed and in section 5 we present our calculations of the  QNMs spectrum and Love numbers through the AdS/CFT duality. In Section 6 we derive the same QNMs spectrum through group theory arguments.  Finally, section 7 contains our conclusions, while Appendix A provide some discussion of the near-zone geometry. In the following we will work with Newton's constant set to unity.

\section{The Kerr spacetime geometry}
The Kerr spacetime describing a rotating BH with mass $M$ and spin parameter  $a$ is a Petrov Type D spacetime as it admits two double principal null directions. 
In the so-called  Boyer-Lindquist coordinates the metric is explicitly written as 
\begin{align}
\dd s^{2} &=  \left(1 - \frac{2Mr}{\Sigma}\right) \dd t^{2} + \frac{4Mra\sin^{2}\theta}{\Sigma}\dd t \dd\phi \nonumber \\
&
- \frac{\Sigma}{\Delta} \dd r^{2} - \Sigma \dd\theta^{2} - \left(r^{2} + a^{2} + \frac{2Mra^{2}}{\Sigma}\sin^{2}\theta\right) \sin^2\theta\dd\phi^{2},
\end{align}
where 

\begin{align}
\Delta &= r^{2} - 2Mr + a^{2},\nonumber\\
\Sigma &= r^{2} + a^{2} \cos^{2}\theta.
\end{align}
The Kerr spacetime has horizons located at  

\be
r_\pm=M\pm \sqrt{M^2-a^2}
\ee
and the corresponding Hawking temperature reads

\be
T_H=\frac{r_+-r_-}{8\pi M r_+}.
\ee
In order to describe the perturbations of the Kerr background, it is standard to introduce the Newman-Penrose formalism \cite{NP}
and we will adopt the  Kinnersley tetrad~\cite{k}
\begin{align}
\ell^{\mu} &= \frac{1}{\Delta} \left( r^{2} + a^{2}, \Delta, 0 , a\right),
\\
n^{\mu} &= \frac{1}{2\Sigma} \left(r^{2} + a^{2}, - \Delta, 0, a\right),
\\
m^{\mu} &= \frac{1}{\sqrt{2}}\,  \frac{1}{r + i a \cos\theta} \left(i a \sin\theta, 0, 1, i \csc\theta\right).
\end{align}
Notice that there are two  vectors,  $\ell^{\mu}$ and $n^{\mu}$,  which are
parallel to the principal null directions of the Kerr background. Because of that, at zeroth order the standard Weyl scalars  are \cite{mm}

\begin{align}
\label{type-D}
   \Psi_{0} = \Psi_{1} = \Psi_{3} = \Psi_{4} 
   = 0,  \quad
   \Psi_{2} =  -\frac{M}{(r-ia\cos\theta)^3}M. 
   \end{align}
At the linear order, the perturbations of the two independent gauge-invariant Weyl scalars obey the 
 Teukolsky equations  \cite{t}

\begin{eqnarray}
\Big[({D}-3\epsilon+\bar{\epsilon}-4\rho-\bar{\rho})(\Delta_n-4\gamma+\mu)-(\delta-3\beta-4\tau+\bar{\pi}-\bar{\alpha})(\bar{\delta}+\pi-4\alpha)-3\Psi_2\Big]\Psi^L_0=0,\nonumber\\
 \label{T1}
\Big[({D}-3\gamma-\bar{\gamma}+4\mu+\bar{\mu})({D}-\rho+4\epsilon)-(\bar{\delta}+\bar{\beta}+3\alpha -\bar{\tau} +4\pi)(\delta-\tau+4\beta) -3\Psi_2\Big]\Psi^L_4=0,\nonumber\\
D=\ell^\mu\partial_\mu,\,\,\Delta_n=n^\mu\partial_\mu, \,\,\delta=m^\mu\partial_\mu, \,\,\bar\delta=\bar m^\mu\partial_\mu.
\label{T2}\nonumber\\
&&
\end{eqnarray}
Due to the  to Goldberg-Sachs theorem, we have  
$ \kappa = \sigma = \lambda = \nu 
   = 0$ and,  in the Kinnersley tetrad, $\epsilon =0$ while  the non-vanishing   spin coefficients turn out to be
   \begin{align}
   \rho &= - \frac{1}{r - i a \cos\theta},
   \qquad
   \beta = -\frac{\bar{\rho}\cot\theta}{2 \sqrt{2} },
   \qquad
   \pi = -\frac{i \bar{\rho}^2 a \sin\theta}{\sqrt{2} },
   \nonumber \\
   \tau &= - \frac{i\rho\bar{\rho}a\sin\theta}{\sqrt{2} },
   \qquad
   \mu =  \frac{1}{2}\, \rho\bar{\rho}^2\Delta ,
   \qquad
   \gamma = \mu + \frac{1}{2}\rho\bar{\rho}(r - M),
   \nonumber \\
   \alpha &= \pi - \bar{\beta}.
 \label{bac}
\end{align}
 Using the explicit expression of Eqs. (\ref{bac}), it turns out that the perturbations 
satisfy \cite{t}
\begin{align}
&\Bigg\{\left[\frac{(r^{2}+a^{2})}{\Delta} - a^{2} \sin^{2}\theta\right] \frac{\partial^{2}}{\partial t^{2}} + \frac{4Mar}{\Delta} \frac{\partial^{2}}{\partial t \partial \phi} - 4 \left[r + i a \cos\theta - \frac{M (r^{2} + a^{2})}{\Delta}\right] \frac{\partial}{\partial t} 
\nn \\
&- \Delta^{-s} \frac{\partial}{\partial r} \left(\Delta^{s+1} \frac{\partial}{\partial r}\right) - \frac{1}{\sin\theta} \frac{\partial}{\partial \theta} \left(\sin\theta \frac{\partial}{\partial\theta}\right) - \left(\frac{1}{\sin^{2}\theta} - \frac{a^{2}}{\Delta}\right) \frac{\partial^{2}}{\partial \phi^{2}} 
\nn \\
&+ 2s \left[\frac{a(r-M)}{\Delta} + \frac{i \cos\theta}{\sin^{2}\theta}\right] \frac{\partial}{\partial \phi} + \left(s^2 \cot^{2}\theta -s\right)\Bigg\}\Psi^L_s = 0,
\end{align}
where $s=0,\pm1,\pm2$ for scalar, vector and tensor perturbations, respectively. The perturbation $\Psi^L_s$ depends on the spin $s$: for $s=0$,   $\Psi^L_0$ is a single scalar (we will call it in the following simply $\Psi$), for $s=+2$, $\Psi^L_{+2}= \Psi^L_0$ and  for $s=-2$,  $\Psi^L_{-2}=\Psi^L_4/\rho^4$. As well known, the   Teukolsky equations are separable. Writing 
\begin{eqnarray}
\Psi^L_s(t,r,\theta,\phi)=e^{-i\omega t}e^{im\phi}S(\theta)R(r), 
\end{eqnarray}
the functions  $R(r)$ and $S(\theta)$ satisfy the following equations
\begin{align}
\label{teuk}
&\Delta^{-s} \frac{\dd}{\dd r} \left(\Delta^{s+1} \frac{\dd R}{\dd r}\right) + \left(\frac{K^{2} - is(\dd \Delta/\dd r)K}{\Delta} +4 is \omega r - \lambda_\omega\right)R = 0,
\\
\label{spher}
&\frac{1}{\sin{\theta}}\frac{\dd}{\dd\theta}\left(\sin{\theta} \frac{\dd S}{\dd\theta}\right) \!+ \!\left(a^{2} \omega^{2} \cos^{2}{\theta} - \frac{m^{2}}{\sin^{2}{\theta}} -2 a \omega s \cos{\theta} \right.\nonumber \\
&\hspace{4cm}\left.-\frac{2 m\,s \cos{\theta}}{\sin^{2}{\theta}} - s^2 \cot^{2}{\theta}  +s+ A\right) S =0,
\end{align}
where 
\begin{eqnarray}
K &=& (r^{2} + a^{2}) \omega - a m,
\nonumber\\
\lambda_{\omega} &=& A+ a^{2} \omega^{2} - 2 a m \omega,  
\label{lambdao}
\end{eqnarray}
and  $ A$ is a separation constant. The solutions to  Eq.~\eqref{spher} are the spin-weighted spheroidal harmonics which in the small $a \omega$ limit reduce to spin-weighted spherical harmonics \cite{g}.

\section{The near-zone region and the ${\rm AdS}_2\times S^2$ geometry}
The goal of this section is to define the so-called near zone region and to show that it can be described by an effective  ${\rm AdS}_2\times S^2$ spacetime geometry for any spin of the possible linear perturbations of the Kerr geometry.

The  near-zone region is  defined for those perturbations satisfying the limit  
\begin{eqnarray}
\omega a\leq \omega r_+<\omega r\ll 1,
\end{eqnarray}
that is we consider perturbations whose Compton wavelength is much larger than the size of the BH. Notice as well that this allows to treat as well the extreme KH BHs for which $a\simeq M$.
Following Ref. \cite{Hui}, we  take 

\be
\frac{r^4}{\Delta(r)}\simeq \frac{r_+^4}{\Delta(r)}
\ee
such that the correct singularity in $r_+$ is preserved and  the  dynamics at larger $r$ is captured.  
  Eqs. (\ref{teuk}) and (\ref{spher}) simplify to 
\begin{align}
\label{teuk1}
&\Delta^{-s} \frac{\dd}{\dd r} \left(\Delta^{s+1} \frac{\dd R}{\dd r}\right) + \left[\frac{(\omega \left(r_+^2 + a^2 \right) -ma)^2 - 2is(r_+ -M)(\omega \left(r_+^2 + a^2 \right)-ma)}{\Delta}\right.\nonumber\\
&-(\ell-s)(\ell+s+1)\Big] R = 0,
\\
\label{spher1}
&\frac{1}{\sin{\theta}}\frac{\dd}{\dd\theta}\left(\sin{\theta} \frac{\dd S}{\dd\theta}\right) + \left(- \frac{m^{2}}{\sin^{2}{\theta}}  -\frac{2 m\,s \cos{\theta}}{\sin^{2}{\theta}} - s^2 \cot^{2}{\theta}  +s+ A_{\ell s}\right) S =0, 
\end{align}
where we have used  the approximations \cite{MS}
\begin{eqnarray}
K^2-\lambda_\omega \Delta &\approx& \left(\omega (r_+^2+ a^2)-m a\right)^2-(\ell-s)(\ell+s+1) \Delta,\nonumber\\
A_{\ell s}&=&(\ell-s)(\ell+s+1)+\mathcal{O}(a^2\omega^2),
\end{eqnarray}
since Eq. (\ref{spher1}) is the equation for spin $s$-spherical harmonics.

For a massless  scalar degree of freedom, the corresponding equation in the near-zone limit can be reproduced by starting from an effective  ${\rm AdS}_2\times S^2$ spacetime geometry  \cite{Hui}
\begin{eqnarray}
\dd s^2&=&\frac{\Delta-a^2 \sin^2\theta}{\r ^2}\dd t^2
+2a \sin^2\theta \dd t 
\dd\phi-\frac{\r ^2}{\Delta}\dd r^2-\r ^2\dd \theta^2-\r ^2 \sin^2\theta \dd\phi^2\nonumber \\
&=&
\frac{\Delta}{\r ^2}\dd t^2-\frac{\r ^2}{\Delta}\dd r^2-\r ^2\left\{\dd \theta^2+ \sin^2\theta \Big(\dd\phi-\frac{a}{\r^2}\dd t\Big)^2\right\}
\label{em}
\end{eqnarray}
where 

\be\r^2=2Mr_+.
\ee
For the interested reader, in Appendix A 
we will demonstrate that indeed this
metric is  an ${\rm AdS}_2\times S^2$ metric and provide some useful alternative forms in different coordinate systems. 
Let us just notice that $r=r_+$ is the horizon location and the AdS$_2$ boundary is located at  $r\gg r_+$.

\subsection{The near-zone region and the ${\rm AdS}_2\times S^2$ geometry for any linear perturbation spin field}
We will now show that the Teukolsky equations on the metric (\ref{em}) are indeed identical to the equations (\ref{teuk1}) and (\ref{spher1}) for any spin $s$. 
 Choosing the null tetrads as 
 \begin{align}
 \ell^\mu&=\frac{1}{\Delta}\Big(\r ^2,\Delta,0,a\Big), \nonumber \\
 n^\mu&=\frac{1}{2\r ^2}\Big(\r ^2,-\Delta,0,a\Big), \nonumber \\
m^\mu&=\frac{1}{\sqrt{2}}\frac{1}{\r }\Big(0,0,1,\frac{i}{\sin\theta}\Big),
 \end{align}
 we find
 \begin{align}
 &\rho=\sigma=\kappa=\tau=\epsilon=\nu=\lambda=\mu=\pi=0, \nonumber \\
 &\alpha=-\frac{1}{2\sqrt{2}}\frac{\cot\theta}{\r }, ~~~~~
 \beta=-\alpha, ~~~~~\gamma=\frac{M-r}{2\r ^2}
 \end{align}
 and the zeroth order Weyl scalars to be 
 \begin{eqnarray}
 \Psi_0=\Psi_1=\Psi_2=\Psi_3=\Psi_4=0.
 \end{eqnarray}
 Therefore the spacetime is of type O in the Petrov classification and conformally flat. Similarly, we get
 the Teukolsky equations (\ref{T1})  for  solutions of the form 
\begin{eqnarray}
\Psi_s^L(t,r,\theta,\phi)=e^{-i\omega t}e^{im\phi}S(\theta)R(r), 
\end{eqnarray}
which  turn out to be in the background  of Eq. (\ref{em}) 
 \begin{align}
\label{teukn}
&\Delta^{-s} \frac{\dd}{\dd r} \left(\Delta^{s+1} \frac{\dd R}{\dd r}\right) + \left(\frac{K^{2} - is(\dd \Delta/\dd r)K}{\Delta} - \lambda_\omega\right)R = 0,
\\
\label{sphern}
&\frac{1}{\sin{\theta}}\frac{\dd}{\dd\theta}\left(\sin{\theta} \frac{\dd S}{\dd\theta}\right) -\left(  \frac{m^{2}}{\sin^{2}{\theta}} +\frac{2 m\,s \cos{\theta}}{\sin^{2}{\theta}} + s^2 \cot^{2}{\theta}  -s- A\right) S=0, 
\end{align}
where now
\begin{eqnarray}
K= \r ^2\omega-m a, \qquad \lambda_{\omega} = A = (\ell-s)(\ell+s+1),
\end{eqnarray}
 since Eq. (\ref{sphern}) 
is  the equation for spherical harmonics. 

Since $\r^2=r_+^2+a^2$, it is easy to verify that Eqs. (\ref{teukn}) and (\ref{sphern}) are identical to the Eqs. (\ref{teuk1}) and (\ref{spher1}) for perturbations in the near-zone of  Kerr background. Therefore, we have shown the AdS$_2\times S_2$ metric (\ref{em}) provides an effective background for the propagation of scalar, vector and tensor perturbations in the near-zone of a Kerr background. 

 \section{The horizon and the ergosphere of the ${\rm AdS}_2\times S^2$ geometry }
In this section  we will show that the basic   properties of the Kerr spacetime can be attributed to the near-zone ${\rm AdS}_2\times S^2$ geometry. In particular, we will show that the latter has, like the Kerr spacetime, an horizon and an ergoshere. In this sense the Kerr BH can be described in terms of the simpler ${\rm AdS}_2$ BH. 

\subsection{Horizon}
We start by defining new coordinates 
\begin{eqnarray}
v=t+\int^r\dd r'\frac{\r ^2 }{\Delta(r')}, ~~~~~
\chi=\phi-\int^r \dd r'\frac{a}{\Delta(r')},
\end{eqnarray}
 thanks to which the metric (\ref{em}) assumes the form 
 \begin{eqnarray}
\dd s^2=\frac{\Delta-a^2 \sin^2\theta}{\r ^2}\dd v^2-2\dd v \dd r
+2a \sin^2\theta \dd v\dd\chi -\r ^2\dd \theta^2-\r ^2 \sin^2\theta \dd\chi^2.  \label{em1}
\end{eqnarray}
 The vector 
 \begin{eqnarray}
 {\bf n}=\frac{\partial}{\partial v}+\frac{\Delta}{\r ^2}\frac{\partial}{\partial r}+\frac{a}{\r ^2}\frac{\partial}{\partial \chi}
 \end{eqnarray}
 is normal to the $r=\mbox{constant}$
surfaces with 
\begin{eqnarray}
n^\mu n_\mu=-\frac{\Delta}{\r ^2}. 
\end{eqnarray}
This implies that $n^\mu$ is null on $r=r_+ $ since $\Delta\Big|_{r_+ }=0$.
In particular,  
\begin{eqnarray}
\left.{\bf n}\right|_{r_+ }=\frac{\partial}{\partial v}+\frac{a}{\r ^2}\frac{\partial}{\partial \chi}\equiv\boldsymbol{\xi}
\end{eqnarray}
 and it is easy to see that $\xi^\mu$ is a Killing vector which generates simultaneous translations in the  $v$ and $\chi$ directions.  
 In addition, in order to find the surface gravity $\kappa$, we need to calculate the eigenvector equation
 \begin{eqnarray}
 \xi^\mu\nabla_\mu \xi^\nu\Big|_{r_+ }=\kappa \xi^\nu .\label{kk}
 \end{eqnarray}
  In other words, the hypersurface $r = r_+ $ is a  Killing horizon of the Killing vector fields
\begin{eqnarray}
\xi^\mu=\left(1,0,0,\frac{a}{\r ^2}\right)
\end{eqnarray}
 and, after evaluating Eq. (\ref{kk}), the  surface gravity is
 \begin{eqnarray}
 \kappa=\frac{\sqrt{M^2-a^2}}{\r ^2}=\frac{r_+-r_-}{2\r^2}.
 \end{eqnarray}
Recalling that the Hawking temperature is 
\begin{eqnarray}
T_H=\frac{\kappa}{2\pi},
\end{eqnarray}
we get 
\begin{eqnarray}
T_H=\frac{r_+-r_-}{8\pi Mr_+}, \label{HTd}
\end{eqnarray}
which is precisely the  Hawking temperature of the Kerr BH. Therefore, 
the effective near-zone geometry has the same Hawking temperature  and shares the same thermodynamic properties with the Kerr geometry.

\subsection{Ergosphere}
The Kerr near-zone specetime with metric  (\ref{em}) or (\ref{em1})
has also another property, namely it has an ergosphere. We are going to show that it is  the same ergosphere of the exact Kerr spacetime. The  vector $\mathbf{k}=\partial/\partial t$ has length 
\begin{eqnarray}
k^2=k^\mu k_\mu=g_{tt}=\frac{\Delta-a^2\sin^2\theta}{\r ^2}
\end{eqnarray}
and is  timelike as long as $k^2>0$ and spacelike for $k^2<0$. In other words, $k^\mu$ is timelike only in the region
\begin{eqnarray}
r^2-2M r+a^2 \cos^2\theta>0. 
\end{eqnarray}
whose boundary 
$r^2-2M r+a^2 \cos^2\theta=0$
defines the ergosphere. The latter touches the event horizon at $\theta = 0$ and $\theta=\pi$, and  it lies outside the horizon for other values of $\theta$. As announced, it precisely coincides with the ergosphere of the Kerr spacetime.

\section{The QNMs and the Love numbers of the Kerr BHs from the AdS$_{\mathbf 2}$ /CFT$_1$ correspondence }
Given the description of the Kerr geometry in the near-zone limit by that of an AdS$_2$ BH, it is natural to conjecture that there is an   ${\rm AdS}_2$/CFT$_1$ duality which  captures the near-zone region physics of the Kerr BH. This correspondence has not to be confused with the so-called Kerr/CFT correspondence valid in the near horizon limit and for extreme Kerr BHs  \cite{kcft}. 

The AdS/CFT correspondence \cite{maldacena} in general states that for every bulk field $\Phi$  there is a corresponding boundary local operator ${\cal O}$ coupled to the boundary field $\Phi_b$ such that \cite{ads1,ads2}

\be
Z_{\rm eff}[\Phi]=e^{iS_{\rm eff}[\Phi]}=\Big<T e^{\int_{\rm boundary}\Phi_b{\cal O}}\Big>,
\ee
The QNMs for BHs in asymptotically AdS spacetimes are defined as the solutions to the wave equation obeying  incoming wave boundary conditions at the horizon and the vanishing Dirichlet condition at the boundary \cite{son,sta}.
In other words, the latter defines the QNM spectrum for which has the interpretation of the poles of the retarded correlators of the operator ${\cal O}$ in the holographically dual theory. In the language of 
linear response theory, the QNMs characterize the time scale needed to approach 
equilibrium in the boundary field theory perturbed by
the operator ${\cal O}$.
Indeed, it has been suggested  that the QNM frequencies are related to the process of
thermalization in the dual strongly coupled CFT \cite{ht}.

Let us  consider   a massless complex scalar field in the background geometry (\ref{em}). Its dynamics is described by the action
\begin{align}
S&=\int \dd^4x\sqrt{-g}\partial_\mu \Phi^*\partial^\mu \Phi\nonumber \\&=\int \dd t \dd r\dd\Omega_2\,  \left(\frac{\r^4}{\Delta}
\Big|\partial_t\Phi+\frac{a}{\r^2}\partial_\phi\Phi\Big|^2
-\Delta \big|\partial_r\Phi\big|^2-\big|\partial_\theta \Phi\big|^2-\frac{1}{\sin^2\theta}\Big|\partial_\phi \Phi\big|^2\right), \label{ac11}
\end{align}
where $\dd\Omega_2$ is the volume element of the two-sphere. 
Expanding the scalar field as 
\begin{eqnarray}
\Phi(t,r,\theta,\phi)=\Psi(t,r) Y_{\ell m}(\theta,\phi)
\end{eqnarray}
where $Y_{\ell m}(\theta,\phi)$ are the spherical harmonics, the action (\ref{ac11}) is written as 
\begin{align}
S&=\int \dd t \dd r  \left(\frac{\r^4}{\Delta}
\big|\big(\partial_t+im\Omega\big)\Psi\big|^2
-\Delta \big|\partial_r\Psi\big|^2-\ell(\ell+1)\big|\Psi\big|^2\right). \label{ac12}
\end{align}
We may further expand the function  $\Psi(t,r)$  as 
\begin{eqnarray}
\Psi(t,r)=\int_{-\infty}^\infty \dd\omega e^{-i\omega t} \Psi(\omega,r).
\end{eqnarray}
In general, suppose that the field  $\Psi(\omega,r)$  which satisfies  the incoming wave boundary condition at the horizon can be written in the basis of two local solutions at the boundary as 

\be
\Psi(\omega,r)=A(\omega)\phi_1(r)+B(\omega)\phi_2(r),
\ee 
where the coefficients  $A$ and  $B$  depend on the frequency $\omega$  which enter the differential equation for the function $\Psi(\omega,r)$.
Near the boundary, one can write 
\begin{align}
\Psi(\omega,r)&=Ar^{-\Delta_-}(1+\cdots)+Br^{-\Delta_+}(1+\cdots)\nonumber \\
&=(r_+-r_-)\Big[A(1-z)^{\Delta_-}(1+\cdots)+B(1-z)^{\Delta_+}(1+\cdots)\Big], \label{dd1}
\end{align}
where $\Delta_{\pm}$ are exponents of the differential equation at the boundary. In such a case, the boundary field is coupled to the operator ${\cal O}$ whose retarded Green function at the boundary (where the field $\Psi(\omega,r)$ is appropriately normalized to unity) $G_R(\omega)$ is proportional to

\be
G_R(\omega)\sim \Psi^*(\omega,r)\partial_r \Psi(\omega,r)|_{\rm boundary}\sim (\Delta_+-\Delta_-)\frac{B}{A}. \label{GR}
\ee
The poles of the retarded correlator therefore correspond to zeros of the  coefficient $A$, which  is precisely the vanishing Dirichlet boundary condition at the boundary  defining the QNMs in asymptotically AdS spacetimes \cite{son}.
For a linear perturbation of spin $s$  in the near-zone ${\rm AdS}_2\times S^2$ region,  the radial equation (\ref{teukn})
turns out to be

\begin{eqnarray}
\Delta^{-s}\partial_r \Big(\Delta^{s+1} \partial_r R^s_{\rm nz}(r)\Big)+\left(\frac{\r^4( \omega-m \Omega)^2-i s\r^2 (\omega-m\Omega)\Delta' }{\Delta}-(\ell-s) (\ell+s+1)\right)R^s_{\rm nz}(r)=0,\nonumber \\
\label{eq22s}
\end{eqnarray}
where

\be
\Omega=\frac{a}{\r^2}.
\ee
By defining the new coordinate 
\begin{eqnarray}
z=\frac{r-r_+}{r-r_-}, \qquad 0\leq z\leq 1, \label{z}
\end{eqnarray}
for which the boundary is located at $z=1$, 
  the equation above can be  written as  

\begin{eqnarray}
&&(1-z)z \partial_z^2 R^s_{\rm nz}+\Big(1+s -(1-s) z\Big)
\partial_z R ^s_{\rm nz}\nonumber \\
&&+\left[\frac{1}{z}\left((1-z)\frac{\xi^2}{4}-i  (1+z)\frac{\xi}{2} s\right)-\frac{(\ell-s)(\ell+s+1)}{1-z}\right]R^s_{\rm nz}=0,
\end{eqnarray}
where we have defined  
\begin{eqnarray}
\xi=\frac{2(ma-\r^2\omega)}{r_+-r_-}=\frac{2\r^2(m\Omega-\omega)}{r_+-r_-}.
\label{rsp}
\end{eqnarray}
The solution with incoming boundary condition $R^s_{\rm nz}(z)\sim z^{i\xi/2-s}$
at $z=0$ ($r=r_+$)  
is given by 

\begin{eqnarray}
R^s_{\rm nz}(z)= C^s_{\rm nz} z^{i\frac{\xi}{2}-s} (1-z)^{\ell+1+s} 
{}_2F_1\left(\ell+1+i \xi,\ell+1-s,1+i\xi-s;z\right),  \label{F212s}
\end{eqnarray}
where $C^s_{\rm nz}$ is a constant. We can now expand the 
solution for general $s$ for large distances within the near-zone limit. By using the modular properties of the hypergeometric functions

\begin{align}
{}_2F_1\left(a,b,c;z\right)&=\frac{\Gamma(c)\Gamma(c-a-b)}{\Gamma(c-a)\Gamma(c-b)}{}_2F_1\left(a,b,a+b-c+1;1-z\right)\nonumber \\
&+
(1-z)^{c-a-b}\frac{\Gamma(c)\Gamma(a+b-c)}{\Gamma(a)\Gamma(b)}{}_2F_1\left(c-a,c-b,c-a-b+1;1-z\right),
\end{align}
we find that for $z\simeq 1$

\begin{eqnarray}
R^s_{\rm nz}(z)&=&C^s_{\rm nz}\Gamma(1-s+i\xi)\bigg[\frac{\Gamma(\Delta_--\Delta_+)}{\Gamma(\Delta_--s)\Gamma(\Delta_-+i\xi)}(1-z)^{\Delta_++s}+\cdots
\nonumber \\
&+&\frac{\Gamma(\Delta_+-\Delta_-)}{\Gamma(\Delta_++i\xi)\Gamma(\Delta_+-s)} (1-z)^{\Delta_-+s}+\cdots\bigg],\label{bss}
\end{eqnarray}
where 
\be
\Delta_+=\ell+1, ~~~~~~\Delta_-=-\ell.
\ee
The  retarded Green function then becomes

\begin{align}
G_R(\omega)&\sim  (\Delta_+-\Delta_-)\frac{B}{A}\sim 
(\Delta_+-\Delta_-)\frac{\Gamma(\Delta_--\Delta_+)}{\Gamma(\Delta_+-\Delta_-)}\frac{\Gamma(\Delta_+-s)}{\Gamma(\Delta_--s)}
\frac{\Gamma(\Delta_++i\xi)}{\Gamma(\Delta_-+i\xi)}.
\label{GRs}
\end{align}

\subsection{ The QNMs of the Kerr BH}
\vskip 0.2cm
\noindent
The poles of the retarded Green function, or alternatively the vanishing Dirichlet boundary condition at the boundary, define
the frequency of the QNMs. The poles are specified by
\be
\Delta_++i\xi=-n, ~~~~~~n\in \mathbb{Z}_+,
\ee
The corresponding QNM spectrum reads
\be
\label{fund}
\omega_{\rm QNM}=m\Omega-i\frac{r_+-r_-}{2\r^2}\left(n+\ell +1\right)=m\Omega-2\pi i T_H \left(n+\ell +1\right) \,\,\, n\in \mathbb{Z}_+,
\ee
consistent with the fact that the retarded Green function can not have poles in the upper half complex $\omega$-plane. The AdS$_2$/CFT$_1$ duality  correctly reproduces 
the fact that, in the extremality limit $a\simeq M$, all modes for $m>0$ cluster at the critical frequency for superradiance, $2 M {\rm Re}\,\omega_{\rm QNM}=m$ and the ${\rm Im}\,\omega_{\rm QNM}$ goes to zero with $T_H$
\cite{br}. Furthermore, it reproduces the know-fact that,  for any spin and for large $n$, the ${\rm Re}\,\omega_{\rm QNM}$ reaches the  asymptote $m\Omega$ \cite{Bertietal}. The spacing $2\pi T_H$ between the ${\rm Im}\,\omega_{\rm QNM}$ \cite{pad,rev,Bertietal} is also reproduced.   Notice that our results are valid for $m>0$ as required for the condition ${\rm Re}\,\omega_{\rm QNM}>0$ imposed by defining incoming waves at the horizon. 

A interesting  step to take would be to start from our results to build up a  perturbation theory through an  expansion around the near-zone enjoying the  AdS$_2$ symmetry  to compute  the spectrum of
the QNMs for small values of $n$ in terms of symmetry breaking
parameters.

\subsection{ The Love numbers of the Kerr BH}
\vskip 0.2cm
\noindent
The AdS$_2$/CFT$_1$ correspondence can be also useful to derive the Love numbers of the Kerr BH \cite{love1}. For of a static external perturbation, the Love number is the coefficient $B/A$ of the subleading term at large distances which goes like

\be
R(r\gg r_+)\simeq A r^{\ell-s}\left[1+\frac{B}{A}
\left(\frac{r}{r_+-r_-}\right)^{-2\ell-1}\right].
\ee
In the non-static case, however, the definition of the Love number is less clear, but by applying the near-zone limit the Love numbers can be read off in the same way as in the static case. In the language of the AdS$_2$/CFT$_1$ correspondence, the Love number $k_L$ is nothing else than coefficient $B/A$, and therefore proportional to the retarded Green function on the boundary. We find that  

\begin{align}
k_L&=\frac{\Gamma(-1-2\ell)\Gamma(\ell+1+i\xi)\Gamma(\ell+1-s)}{\Gamma(2\ell+1)\Gamma(-\ell+i\xi)\Gamma(-\ell-s)}    \nonumber \\ &=(-1)^{\ell+s+1}\frac{(\ell-s)!(\ell+s)!}{2(2\ell)!(2\ell+1)!}\prod_{j=-\ell}^\ell(j+i\xi)
\end{align}
Expanding in  the limit of $\omega\sim m\Omega$, or $\xi\sim 0$, and sing the expressions 

\be
\frac{\psi(-\ell)}{\Gamma(-\ell)}=(-1)^{\ell+1} \ell!, ~~~~
\frac{\Gamma(-2\ell-1)}{\Gamma(-\ell)}=(-1)^{\ell+1}
\frac{(\ell+1)!}{2(2\ell+1)!},
\ee
 we find
that the Love number is purely imaginary (and  called the dissipative coefficient) 

\begin{align}
k_L&\approx- i\frac{\xi}{2} \frac{\Gamma(-1-2\ell)\Gamma(\ell+1)\Gamma(\ell+1-s)\psi(-\ell)}{\Gamma(2\ell+1)\Gamma(-\ell)\Gamma(-\ell-s)}    \nonumber \\ &=i \frac{\omega-m\Omega}{4\pi T_H} (-1)^s\frac{(\ell-s)!(\ell+s)!(\ell!)^2}{(2\ell)!(2\ell+1)!},
\end{align}
which reproduces the result in Ref. \cite{aa}. This is not surprising as non-extremal BHs  correspond  to excited finite temperature states on the CFT side and thermalization and dissipative properties should be readable  in the dual strongly coupled CFT.

\section{The QNMs of the Kerr BHs from group theory}
In this subsection, we offer a simple group theory approach to reproduce the spectrum of the QNMs based on the AdS$_2$ geometry. The metric (\ref{em}) has the following Killing vectors \cite{stro,Hui}

\begin{align}
L_0&=-\frac{4M r_+ }{r_+ -r_-}\partial_t-\frac{2a}{r_+ -r_-}\partial_\phi, \label{kil1}\\
L_+&=e^{-\frac{(r_+ -r_-)t}{4Mr_+ }}\left(\frac{4 M r_+ }{(r_+ -r_-)}\frac{M-r}{\sqrt{\Delta}}\partial_t-\sqrt{\Delta}\partial_r +\frac{2a}{(r_+ -r_-)}\frac{M-r}{\sqrt{\Delta}}\partial_\phi\right), \label{k2}\\
L_-&=e^{\frac{(r_+ -r_-)t}{4Mr_+ }}\left(\frac{4 M r_+ }{(r_+ -r_-)}\frac{M-r}{\sqrt{\Delta}}\partial_t+\sqrt{\Delta}\partial_r +\frac{2a}{(r_+ -r_-)}\frac{M-r}{\sqrt{\Delta}}\partial_\phi\right), \label{k3}\\
J_+&=-ie^{i\left(\phi-\frac{a}{2Mr_+ }t\right)}\left(\partial_\theta+i \cot\theta \partial_\phi\right), \label{ks1}\\
J_-&=-ie^{-i\left(\phi-\frac{a}{2Mr_+ }t\right)}\left(\partial_\theta-i \cot\theta \partial_\phi\right), \label{ks2}\\
J_0&=-i\partial_\phi, \label{ks3}
\end{align}
which satisfy the ${\rm SL}(2,\mathbb{R})\times {\rm SO}(3)$ algebra

\begin{align}
&[L_+,L_-]=-2 L_0, \qquad[L_0,L_\pm]=\pm L_\pm,\label{sl2}\\
&[J_+,J_-]=2J_0,\qquad [J_0,J_\pm]=\pm J_\pm. \label{so3}
\end{align}
Fields on the 4D near-zone geometry will form representations of ${{\rm SL}(2,\mathbb{R})}\times { {\rm SO}(3)}$. The maximum set of commuting operators are 
$L_0,L^2,J_0,J^2$, where

\begin{align}
L^2
&
=\frac{1}{2}\left(L_+L_-+L_-L_+\right)-L_0^2\nonumber \\
&=L_+L_--L_0(L_0-1), \label{cas1}\\
J^2
&=\frac{1}{2}\left(J_+J_-+J_-J_+\right)+J_0^2\nonumber \\
&=J_+J_-+J_0(J_0-1).
\label{cas2}
\end{align}
are  the quadratic Casimirs for ${{\rm SL}(2,\mathbb{R})}$ and ${ {\rm SO}(3)}$, respectively,
Therefore, fields in the near-zone geometry will be
 labeled as $\Phi_{h,E_0,\ell,m}(x)=\langle x|h,E_0;\ell,m\rangle$ where 
$(h,E_0,\ell,m)$ correspond to eigenvalues of the $L_0,C^2_{{\rm SL}(2,\mathbb{R})},C^2_{ {\rm SO}(3)}$ and $J_0$ operators, respectively, 

\begin{eqnarray}
&&L_0 \Phi_{h,E_0,\ell,m}=E_0\Phi_{h,E_0,\ell,m}, \qquad 
L^2
\Phi_{h,E_0,\ell,m}=h(h-1)\Phi_{h,E_0,\ell,m},\nonumber \\
&&
J_0\Phi_{h,E_0,\ell,m}=m\Phi_{h,E_0,\ell,m},\qquad
J^2\Phi_{h,E_0,\ell,m}
=\ell(\ell+1)\Phi_{h,E_0,\ell,m}. \label{lab}
\end{eqnarray}

For a  massless field $\Phi_{h,E_0,\ell,m}$,
 one can verify  that 
\begin{eqnarray}
\nabla^\mu \nabla_\mu \Phi_{h,E_0,\ell,m}=-\Big(L^2
+J^2
\Big)\Phi_{h,E_0,\ell,m}=0. 
\label{na}
\end{eqnarray}
Therefore, we find that 

\be
L^2\Phi_{h,E_0,\ell,m}=-J^2\Phi_{h,E_0,\ell,m},
\ee
which, using Eq. (\ref{lab}) implies that $h=h_\pm$ with

\be
h_+=\ell+1, ~~~~~h_-=-\ell. 
\ee
On the other hand,
 we have  expressed  the modes $\Phi_{h,E_0,\ell,m}$ as 
\begin{eqnarray}
\Phi_{h,\omega,\ell,m}(t,r,\theta,\phi)=e^{-i\omega  t}R_{\omega,m}(r) Y_{\ell m}(\theta,\phi), \label{Phi}
\end{eqnarray}
where $Y_{\ell m}(\theta,\phi)$ are spherical harmonics. 
The lowest (highest) weight mode $\Phi^{(0)}_{h,E_0,\ell,m}$  is defined by $L_-\Phi^{(0)}_{h,E_0,\ell,m}=0$  ($L_+\Phi^{(0)}_{h,E_0,\ell,m}=0$) and since $L_0 \Phi^{(0)}_{h,E_0,\ell,m}=E_0 \Phi^{(0)}_{h,E_0,\ell,m}$, one finds

\begin{eqnarray}
E_0=2i \frac{2Mr_+ \omega-ma}{r_+ -r_-}=-i\xi. \label{e0}
\end{eqnarray}
We are interested in representations with $E_0$ bounded from below (since $E_0$ is related to the energy as it is  eigenvalue of the time translation operator), and therefore, we will consider
fields belonging to the discrete representations ${\cal{D}}^+(h_+,E_0)$ of ${{\rm SL}(2,\mathbb{R})}$ \cite{barg,barut}. For such representations we have 
\be 
E_0=h_++n=\ell+1+n, ~~~~n=0,1,2,\cdots,
\ee
and therefore, using Eq. (\ref{e0}), we get that 
\be
-i\xi=n+\ell+1,
\ee
recovering  once more the QNM spectrum in Eq. (\ref{fund}).

\section{Conclusions}
We have used the near-zone region of the Kerr BH to investigate the spectrum of highly damped QNMs. First, we have shown that scalar, vector and tensor linear degrees of freedom obey in the near-zone limit the same equations derived from an effective
AdS$_2\times S^2$ geometry which enjoys the same basic properties of the Kerr geometry. We have subsequently  made use of the  the AdS$_2$/CFT$_1$ correspondence to obtain the QNMs spectrum and the Love numbers of  Kerr BHs. The same result has been also obtained through a group theory method. 

Our findings can be extended in several ways. For instance, we will investigate the linear instability of the massive Klein-Gordon field in the Kerr geometry by employing the effective description of an AdS$_2$ BH. Furthermore, and possibly, more interestingly, one could   construct  a perturbation theory  by starting  with an expansion around the near-zone, where perturbations at the linear level enjoy the AdS$_2$ symmetry and calculate the spectrum of the QNMs  as a series in terms   of symmetry breaking parameters.  
In such a way one could also study the nonlinearities in Kerr BH ringdowns  which have been recently shown to be relevant \cite{c1,c2,c3}. For instance, it might be possible to understand the relevance of the nonlinearities from group theory arguments, that is from the fact that nonlinearities can be written as a sum of potentially large quadratic linear terms through a Clebsch-Gordan decomposition of the ${{\rm SL}(2,\mathbb{R})}$ group. Alternatively, one could calculate the cubic action of tensor modes from the AdS$_2\times S^2$ geometry using the ADM formalism following the  standard procedure
of Ref. \cite{mng}.

\section*{Acknowledgments}
We thank V. De Luca, K. Kokkotas and P. Pani for useful discussions. The work of D. P.  is supported by the Swiss National Science Foundation under grants no. 200021-205016 and PP00P2-206149. A.K. is supported by the PEVE-2020 NTUA programme for basic research with project number 65228100. A.R. is funded by the Boninchi Foundation.

\appendix
\section{The near-zone ${\rm AdS}_2\times S^2$ geometry } 
\label{apA}
\setcounter{equation}{0}
\renewcommand\theequation{A.\arabic{equation}}

For the benfit of the reader we collect here some coordinate systems \cite{ss} to express the 
${\rm AdS}_2\times S^2$ metric of the
 near-zone region. Starting from the effective metric

\begin{eqnarray}
\dd s^2=\frac{\Delta-a^2 \sin^2\theta}{\r ^2}\dd t^2
+2a \sin^2\theta \dd t 
\dd\phi-\frac{\r ^2}{\Delta}\dd r^2-\r ^2\dd \theta^2-\r ^2 \sin^2\theta \dd\phi^2,  \label{emm}
\end{eqnarray}
upon defining the new coordinates 

\begin{eqnarray}
\rho=\r \ln\left(2\sqrt{\Delta}+2(r-M)\right)-\r \ln(r_+-r_-), \quad \varphi=\phi-at,
\end{eqnarray}
the metric factorizes as 
\begin{eqnarray}
\dd s^2=\dd s_1^2-\r^2\dd\Omega_2^2, \label{ds0}
\end{eqnarray}
where 
\begin{eqnarray}
&&\dd s_1^2=\frac{(r_+-r_-)^2}{4\r^2 }\sinh^2\left(\frac{\rho}{\r
 }\right)\dd t^2-\dd\rho^2,\label{ds1}\\
 &&\dd\Omega_2^2=\left(d\theta^2+ \sin^2\theta d\varphi^2\right).  \label{ds2}
\end{eqnarray}
The metric (\ref{ds2}) is the metric of the unit $S^2$, whereas, the metric (\ref{ds1}) is the metric on ${\rm AdS}_2$. 
Close to $\rho=0$, the metric (\ref{ds0}) behaves as 
\begin{eqnarray}
\dd s^2\approx \frac{(r_+-r_-)^2}{4\r^4 }\rho^2\dd t^2-\dd\rho^2-\r^2\dd\Omega_2^2.\label{ds12}
\end{eqnarray}
After rotating to Euclidean time $t_E$ and defining 
$\tau_E=t_E (r_+-r_-)/2\r^2 $, 
the metric  (\ref{ds12})   is the flat two-dimensional Euclidean metric in polar coordinates (plus the metric on $S^2$)
\begin{eqnarray}
\dd s^2\approx -\rho^2\dd\tau_E^2-\dd\rho^2-\r^2\dd\Omega_2^2 \label{ds11}
\end{eqnarray}
provided the periodicity of $\tau_E$ is $2\pi$. Therefore the periodicity of $t_E$ is $\beta=\frac{4\pi\r^2}{r_+-r_-}$
which corresponds to Hawking temperature 
\begin{eqnarray}
 T_H=\beta^{-1}=\frac{r_+-r_-}{4\pi\r^2}. \label{Hdd}
 \end{eqnarray} 
If we recall that $\r^2=2M r_+$,  the Hawking temperature 
(\ref{Hdd}) is exactly the Hawking temperature of the Kerr black hole as we find also in (\ref{HTd}). 
 We can also write the metric (\ref{em}) in  conformal coordinates 
 $(\tau,x)$ as 
 \begin{eqnarray}
 \dd s^2=\frac{(r_+-r_-)^2}{4\r^2 }{\rm csch}^2\left(\frac{r_+-r_-}{2\r}x\right) \Big(\dd\tau^2-dx^2\Big)-\r^2\dd\Omega_2^2,\label{ds22}
 \end{eqnarray}
 where
 \begin{eqnarray}
 \tau=t,\qquad x=\frac{\r^2}{r_+-r_-}\ln\frac{r-r_+}{r-r_-}.
 \end{eqnarray}
Furthermore, by defining new  coordinates $(\tau,\sigma)$ by 
\begin{eqnarray}
\tau\pm\sigma\pm\frac{\pi}{2}=2\tan^{-1}\left\{
\tanh\left[\frac{r_+-r_-}{4 \r}
\left(t\pm\frac{\r^2}{r_+-r_-}\ln\frac{r-r_+}{r-r_-}    \right)\right]
\right\},
\end{eqnarray}
the metric (\ref{em}) is written as 
\begin{eqnarray}
\dd s^2=\r^2\left[\frac{1}{\cos^2\sigma}(\dd\tau^2-\dd\sigma^2)-\dd\Omega_2^2\right].
\end{eqnarray}
Finally, defining

\be
T\pm y=\tan\frac{1}{2}(\tau\pm \sigma\pm \pi/2),
\ee
one recovers the familiar metric

\be
\dd s^2=\r^2\left[\frac{1}{y^2}(\dd T^2-\dd y^2)-\dd\Omega_2^2
\right],
\ee
where the horizon is at $y\to \infty$.



\end{document}